# A Spontaneously Formed Plasmonic–MoTe$_2$ Hybrid Platform for Ultrasensitive Raman Enhancement

Li Tao[1,2]*, Zhiyong Li[1], Kun Chen[3]*, Yaoqiang Zhou[1], Hao Li[1], Ximiao Wang[3], Runze Zhan[3], Xiangyu Hou[4], Yu Zhao[5], Junling Xu[6], Teng Qiu[4], Xi Wan[7], Jian-Bin Xu[1]*

1. Department of Electronic Engineering, The Chinese University of Hong Kong, Hong Kong SAR, China

2. Key Lab of Advanced Optoelectronic Quantum Architecture and Measurement (Ministry of Education), School of Physics, Beijing Institute of Technology, Beijing 100081, China

3. State Key Laboratory of Optoelectronic Materials and Technologies, School of Electronics and Information Technology and Guangdong Province Key Laboratory of Display Material, Sun Yat-sen University, Guangzhou 510275, China

4. School of Physics, Southeast University, Nanjing 211189, China

5. Guangdong Provincial Key Laboratory of Functional Soft Condensed Matter, School of Material and Energy, Guangdong University of Technology, Guangzhou 510006, China

6. Institute of Textiles and Clothing, The Hong Kong Polytechnic University, Hong Kong SAR, China

7. Department of Electronic Engineering, Jiangnan University, Wuxi 214122, China

*Corresponding authors. Emails: taoli@link.cuhk.edu.hk (L.T.); chenk69@mail.sysu.edu.cn (K.C.); jbxu@ee.cuhk.edu.hk (J.-B.X.)

**SUMMARY**

To develop highly sensitive, stable and repeatable surface-enhanced Raman scattering (SERS) substrates is crucial for analytical detection, which is a challenge for traditional metallic structures. Herein, by taking advantage of the high surface activity of 1T′ transition metal telluride, we have fabricated high-density gold nanoparticles (AuNPs) that are spontaneously in-situ prepared on the 1T′ MoTe$_2$ atomic layers via a facile method, forming a plasmonic–2D material hybrid SERS substrate. This AuNP formation is unique to the 1T′ phase, which is repressed in 2H MoTe$_2$ with less surface activity. The hybrid structure generates coupling effects of electromagnetic and chemical enhancements, as well as excellent molecule adsorption, leading to the ultrasensitive (4×10$^{-17}$ M) and reproducible detection. Additionally, the immense fluorescence and photobleaching phenomena are mostly avoided. Flexible SERS tapes have been demonstrated in practical applications. Our approach facilitates the ultrasensitive SERS detection by a facile method, as well as the better mechanistic understanding of SERS beyond plasmonic effects.

**INTRODUCTION**

Surface-enhanced Raman scattering (SERS) has been intensively investigated as a promising analytical tool for non-destructive and label-free fingerprint molecular detection.[1] Metallic plasmonic nanostructures are widely studied as sensitive SERS platforms owing to the largely augmented electromagnetic field generated by localized surface plasmon resonance (LSPR) at the hot spots, known as the electromagnetic mechanism (EM).[2] Yet the noble metal structures for SERS suffer from significant limitations including: 1) complicated and delicate fabrication process with design constraints; 2) low reproducibility and uniformity; 3) large analyte fluorescence (FL) backgrounds; and 4) metal-catalyzed side reactions and photobleaching effect.[3] In recent years, graphene and other 2D layered materials have drawn great attention as a new type of SERS substrates.[4–7] The atomic flat 2D materials uniformly adsorb the analytes, forming strongly interacted interfaces. In contrast with the plasmonic structures, the SERS effect on 2D material substrates relies on the chemical mechanism (CM) that stems from the photo-induced charge transfer in the analyte–2D material interface.[8–10] The CM-based 2D material substrates show advantages over the noble metal SERS structures in terms of stability, reproducibility, quenched FL signals, etc. Recently we have reported novel 1T′ transition metal telluride (TMT) atomic layers as ultrasensitive CM-based SERS platforms.[11] The semi-metallic tellurides generate strong coupling with the adsorbed molecules and provide a high density of states for the charge transfer, thus delivering much better SERS performance over graphene. Other semi-metallic 2D materials with abundant low-energy states such as $NbS_2$ and $ReO_xS_y$ are demonstrated later to exhibit superior SERS sensitivities.[12,13]

Hybrid SERS platforms consisting of plasmonic structures and 2D materials are proposed to be promising candidates for next-generation molecular sensors, because they allow the combination of advantages of both EM and CM.[14] Previous studies have shown the 2D graphene and hexagonal boron nitride (hBN) films hybridized with thermally deposited noble metal nanoparticles can efficiently enhance the SERS sensitivity and partly solve the above-mentioned challenging issues of noble metals.[15–21] The 2D materials serve as a shield for noble metals, provide additional chemical enhancement, and reduce the molecular FL. However, the chemical enhancement of graphene and hBN is limited and the thermal deposition of noble metal nanoparticles involves additional costly fabrication processes. In addition, depositing of noble metal nanoparticles on 2D materials causes potential damage to the 2D materials,[22] while covering pre-deposited noble metal nanoparticles with 2D materials introduces interface quality degradation during the transfer processes.

On the contrary, 2D TMT can effectively enrich the adsorbed probe molecules and generates significant chemical enhancement. More importantly, the low-symmetry atomic structure of 1T′ TMT makes its surface highly active, which ensures the strongly interacted analyte–TMT interface, leading to the coupled electromagnetic and chemical enhancements for SERS in the plasmonic–TMT hybrid structure. In the present study, we report on a facile and high-throughput fabrication of the hybrid structure of $MoTe_2$ atomic layers and gold nanoparticles (AuNPs) as a high-performance SERS mediator. In contrast with thermal deposition, the uniformly distributed AuNPs are spontaneously in-situ fabricated on the $MoTe_2$ surface by redox reaction with the gold precursor. The unique large surface activity of the 1T′ phase $MoTe_2$ plays the key role in the formation of AuNPs, evidenced by comparative investigation on $MoTe_2$ with different phases. The plasmonic–$MoTe_2$ hybrid structure exhibits an ultralow detection limit of $4\times10^{-17}$ M for Rhodamine 6G (R6G) probe, which is four orders of magnitude improved in comparison with the individual $MoTe_2$, and the SERS signals are reproducible, stable, uniform and clean. The practical application including food safety control is demonstrated in the AuNPs–$MoTe_2$ platform, showing great potential in highly sensitive and reliable SERS detection.

**RESULTS**

**Fabrication of the plasmonic–$MoTe_2$ hybrid platform**

Large-scale and highly crystalline trilayer (3L) $MoTe_2$ samples synthesized by salt-assisted chemical vapor deposition (CVD) method were adopted in the experiments, as shown in **Figure 1A**. The highly crystalline and atomic flat features of the as-grown 1T′ $MoTe_2$ were confirmed by Raman spectroscopy, atomic force microscopy (AFM), and high-resolution transmission electron microscopy (HRTEM) depicted in **Figure S1**. The in-situ formation of AuNPs on the 1T′ MoTe$_2$ atomic layers was realized by simply immersing the sample in HAuCl$_4$ solution in ethanol (1 mM) for 30 s (**Figure 1B**), followed by gentle drying with nitrogen flow. The scanning electron microscopy (SEM) images of 1T′ MoTe$_2$ before and after the treatment (**Figure 1C and 1D**) reveal the dense and uniform morphology of AuNPs decorated on the MoTe$_2$ surface with diameters ~20 nm. Importantly, the AuNPs were exclusively found on MoTe$_2$, while the SiO$_2$ regions remained intact. The 2D 1T′ MoTe$_2$ has considerable surface activity associated with the distorted octahedral lattice, leading to the spontaneous in-situ transition from [AuCl$_4$]$^-$ into AuNPs in the HAuCl$_4$ solution without additional reduction agent.[23,24] In contrast, reduction agents such as trisodium citrate or pretreatments such as laser processing are required for growing AuNPs on other less-active 2D materials such as graphene and MoS$_2$ in most synthetic strategies.[25–27] The atomic structure of the AuNPs–MoTe$_2$ hybrid was investigated by HRTEM as shown in **Figure 1E**. The MoTe$_2$ at the nanogaps between the AuNPs exhibited excellent 1T′ octahedral lattice, indicating that the MoTe$_2$ was well preserved after the HAuCl$_4$ treatment. Additional TEM images are available in **Figure S2A and S2B**. **Figure 1F** gives the high-angle annular dark field (HAADF) image of the AuNPs–MoTe$_2$ and the elemental maps collected by energy-dispersive X-ray spectrometer. The Au map fits well with the bright areas in the HAADF image, while the uniformly distributed Mo and Te elements reveal the intact MoTe$_2$ film underneath. X-ray photoelectron spectroscopy (XPS) investigation on the MoTe$_2$ flake shown in **Figure S2C-S2G** further revealed the emerging of Au 4f$_{5/2}$ and Au 4f$_{7/2}$ peaks derived from the AuNPs after HAuCl$_4$ treatment, while the positions and shapes of Te 3d and Mo 3d peaks were unchanged after the treatment.

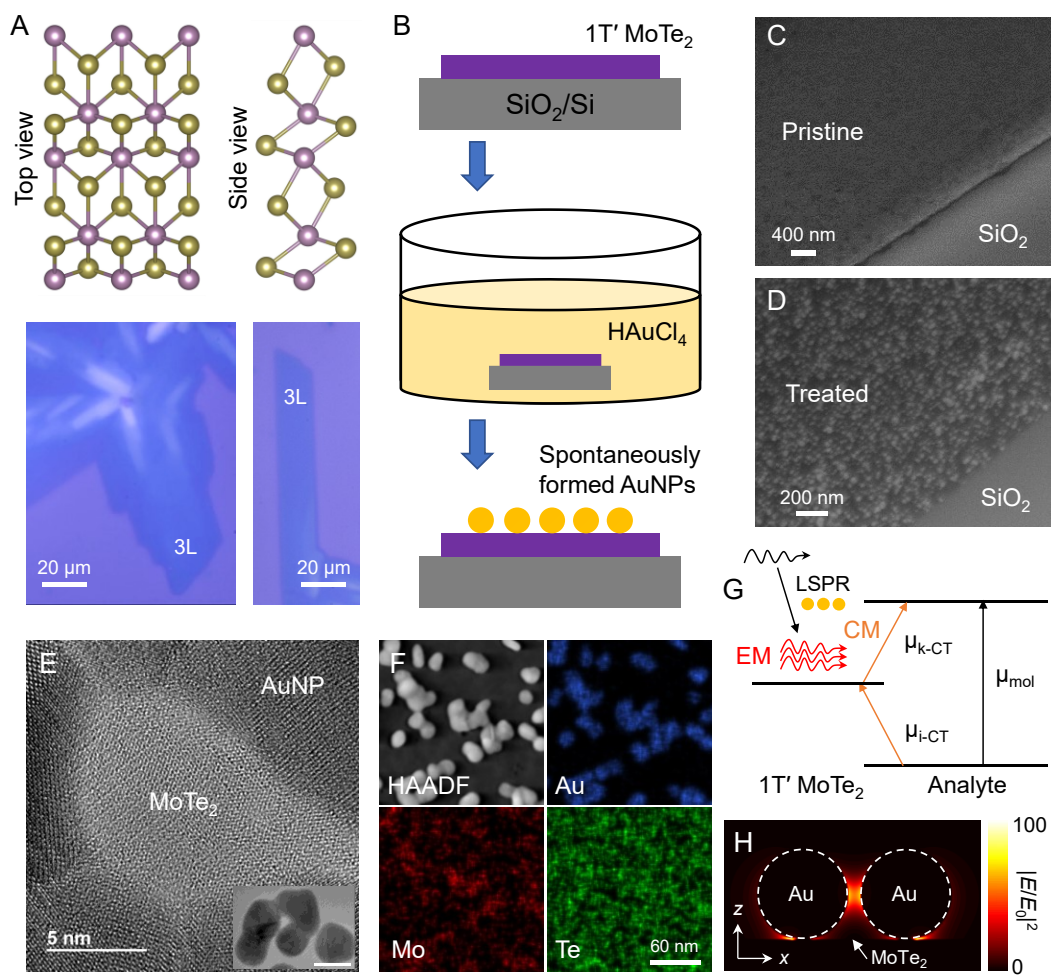

**Figure 1. Spontaneous formation of AuNPs on 1T′ MoTe$_2$ for SERS with EM–CM coupling.**

(A) Lattice structure and optical images of 1T′ MoTe$_2$ atomic layers.

(B) Schematic illustration of the spontaneous in-situ formation of AuNPs on the surface of 1T′ MoTe$_2$.

(C and D) SEM images of the 1T′ MoTe$_2$ before and after the decoration of AuNPs.

(E) HRTEM image of the AuNPs–MoTe$_2$ structure. Inset is a lower magnification image. Scale bar in the inset: 20 nm.

(F) HAADF image and the corresponding elemental maps of the AuNPs–MoTe$_2$ structure.

(G) Illustration showing the coupling of EM and CM in the AuNPs–MoTe$_2$ hybrid SERS structure.

(H) Simulated electrical field distribution of the AuNPs–MoTe$_2$ structure (front view). Diameter of the AuNPs: 20 nm; gap between the AuNPs: 3 nm.

**Coupling of EM and CM in the hybrid platform**

The metallic plasmonic structure creates hot spots where significantly enhanced field intensity lies, but the limited adsorption capability of some probe molecules makes the hot spots in bare noble metallic structures fail to function their full role. 1T′ $MoTe_2$ as a 2D material, on the other hand, allows the excellent surface adsorption of probes because of its low-symmetry 1T′ phase and the π−π interaction at the analyte–$MoTe_2$ interface. As a result, the AuNPs–$MoTe_2$ hybrid delivers a superior platform with increased SERS sensitivity. The $MoTe_2$ not only augments the number of analytes adsorbed at hot spot regions, but also introduces coupled chemical enhancement. The charge transfer resonances between analytes and $MoTe_2$ are largely promoted by leveraging the nearby molecular transition in resonance condition via vibronic coupling process,[28] and can be further enhanced by the strong electromagnetic field generated by the high-density hot spots between AuNPs. In a word, the promising SERS effect of AuNPs–$MoTe_2$ hybrid results from cooperative and coupling effects of plasmon at hot spots, chemical enhancement owing to efficient charge transfer, and enriched molecular adsorption (see the illustration in **Figure 1G** and discussion in **Note S1**).[29] **Figure 1H** depicts the electric field distribution of AuNPs on $MoTe_2$ surface under 532 nm illumination simulated by finite difference time domain (FDTD) method. Highly localized electromagnetic field distributes in the nano-gaps between AuNPs and at the touchpoint between AuNP and $MoTe_2$ sheet, consistent with the analysis above. The AuNPs–$MoTe_2$ hybrid exhibits a LSPR peak around 530 nm, as shown in the extinction spectrum in **Figure S3**.

**Impact of the $MoTe_2$ phase on the AuNP formation**

As discussed above, the spontaneous in-situ formation of AuNPs can be realized on 1T′ MoTe$_2$, while it is not applicable on 2H MoTe$_2$ with low surface activity.[30] SEM and AFM images of the MoTe$_2$ possessing different phases after HAuCl$_4$ treatment are shown in **Figure 1D, 2A-2C**. Further, we have investigated the morphologies of AuNPs on MoTe$_2$ using the dark-field microscopy technique which enables the high-contrast imaging of plasmonic nanoparticles (**Figure 2D and 2E**).[31] The characterization results show that high-density AuNPs are formed on the surface of 1T′ MoTe$_2$, while only few large Au aggregates are detected on 2H MoTe$_2$. As shown in **Figure S4**, the HAuCl$_4$ treatment does not result in measurable improvement for the SERS sensitivity of 2H MoTe$_2$. The significant difference between the morphologies of the HAuCl$_4$ treated 1T′ and 2H MoTe$_2$ can be illustrated by density-functional theory (DFT) calculation, as manifested in **Figure 2F and 2G**. The DFT results show that the energy barrier ($E_b$) for the dechlorinating of [AuCl$_4$]$^-$ ion on 1T′ MoTe$_2$ surface is only 11.23 kcal/mol, while $E_b$ reaches 25.62 kcal/mol on 2H MoTe$_2$ surface. Thus, we can draw the intuitive scenario of reaction rate $k$ for [AuCl$_4$]$^-$ ion dechlorination on 1T′ and 2H MoTe$_2$ surface through a simple unimolecular reaction model:[32]

$$k = \frac{k_B T}{h} \left(\frac{RT}{P_0}\right)^{\Delta n} \exp\left(-\frac{E_b}{k_B T}\right), \qquad (1)$$

where $k_B$ is the Boltzmann constant, $h$ is the Planck constant, $T$ is thermodynamic temperature, $R$ is the universal gas constant and $P_0$ is the standard atmosphere, and for unimolecular reaction $\Delta n=0$. From Eqn. (1), it can be obtained that, at room temperature ($T$=298.15 K), the half-life of [AuCl$_4$]$^-$ ion on the 1T′ MoTe$_2$ surface $t(1/2)=\ln(2/k)$ is much less than 1s. However, on the 2H MoTe$_2$ surface, the half-life of [AuCl$_4$]$^-$ ion is ~190 h. From such remarkable difference of [AuCl$_4$]$^-$ ion half-lives on the 1T′ and 2H MoTe$_2$ surface, we can conclude that, on the surface of 1T′ MoTe$_2$, [AuCl$_4$]$^-$ ion can be fleetly reduced to AuNPs at room temperature, while on the surface of 2H

MoTe$_2$, the dechlorination of [AuCl$_4$]$^-$ ion cannot spontaneously proceed unless on the defective sites. In a word, the 1T′ phase possessing high surface activity largely facilitates the AuNP formation. Thus, in stark contrast with the 2H counterpart, the 1T′ MoTe$_2$ not only itself has considerably larger CM-based SERS sensitivity,[11,33] but also benefits the spontaneous formation of plasmonic nanoparticles, making the AuNPs–MoTe$_2$ hybrid an ultrasensitive and easily fabricated platform for SERS with EM–CM coupling.

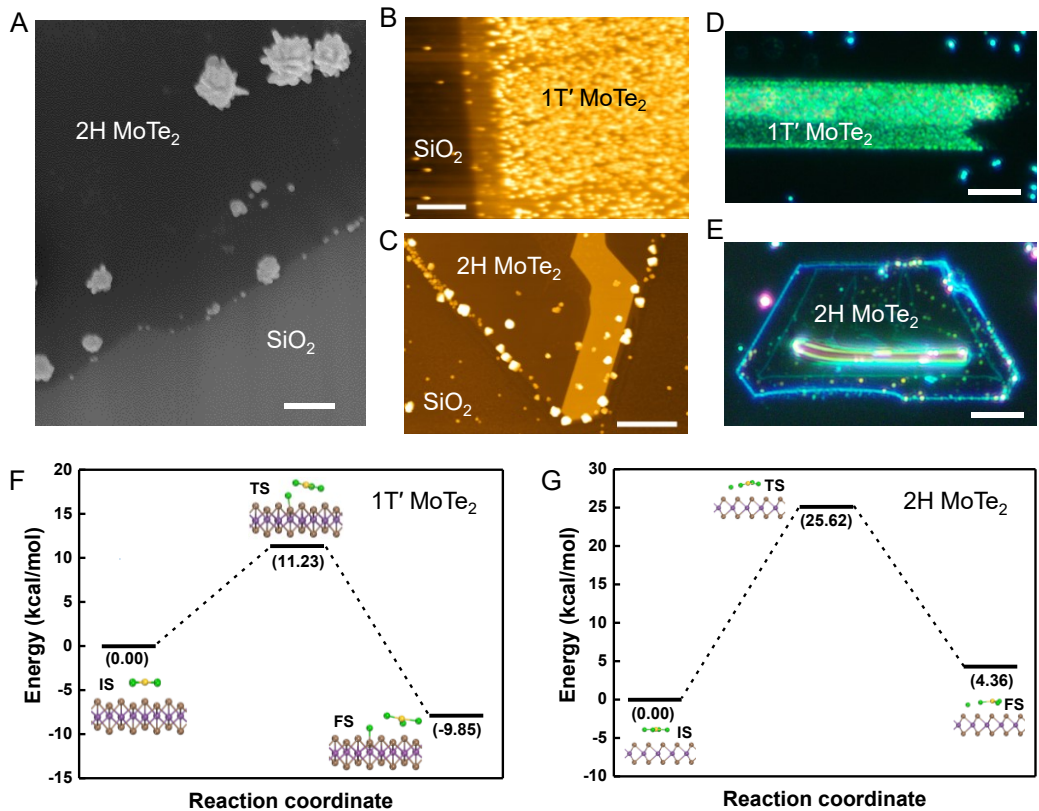

**Figure 2. Formation of AuNPs on 1T′ and 2H MoTe$_2$: the phase matters.**

(A) SEM image of 2H MoTe$_2$ (3L) after HAuCl$_4$ treatment for 30 s. Scale bar: 400 nm.

(B and C) AFM images of 1T′ (B) and 2H (C) MoTe$_2$ after the HAuCl$_4$ treatment. Scale bars: 2 μm.

(D and E) Dark-field optical images of 1T′ (D) and 2H (E) MoTe$_2$ after the HAuCl$_4$ treatment. Scale bars: 10 μm.

(F and G) The minimum energy pathways including the initial state (IS), the transition state (TS), and the final state (FS) for the dechlorinating of [AuCl$_4$]$^-$ ion on 1T′ (F) and 2H (G) MoTe$_2$ surfaces.

**SERS performance of the hybrid platform**

**Figure 3A** shows the direct comparison of the SERS signals (with baseline correction) of a typical probe R6G with 4×10$^{-7}$ M concentration drop-casted on bare 1T′ MoTe$_2$ and AuNPs–MoTe$_2$ hybrid substrates (532 nm excitation). The detected R6G Raman-active peaks on the hybrid structure are largely stronger than those on pure 1T′ MoTe$_2$, though the 1T′ MoTe$_2$ has been demonstrated to present much more sensitive SERS effect than the intensively studied graphene, $h$BN and MoS$_2$.[11] We have also found that the SERS sensitivity of few-layer WTe$_2$ as another 1T′ TMT material is largely enhanced after HAuCl$_4$ treatment, similar with the case of 1T′ MoTe$_2$ (**Figure S5**). Additionally, the 1T′ MoTe$_2$ acts as a promising analyte FL quencher, delivering clean SERS signals, while the large FL background is a major hurdle of resonance SERS on metallic plasmonic substrates.[34,35] As shown in **Figure 3B**, the SERS outputs on AuNPs–MoTe$_2$ are of high signal-to-background ratio, while the signals on bare SiO$_2$ performs large FL background of R6G with no detectable analyte Raman peaks. The FL quenching effect is a consequence of charge transfer and energy transfer between R6G and MoTe$_2$.[11,25] It is worth noting that the formation procedure of dense and self-assembled AuNPs on MoTe$_2$ is simple and efficient within 30 s, and further increase of immersing time in HAuCl$_4$ solution does not result in more effective SERS performance of the fabricated AuNPs–MoTe$_2$ (**Figure S6A and S6B**). This is attributed to the aggregation of AuNPs with prolonged reaction time, which decreases the density of hot spots (see SEM images in **Figure S6C-S6F**).

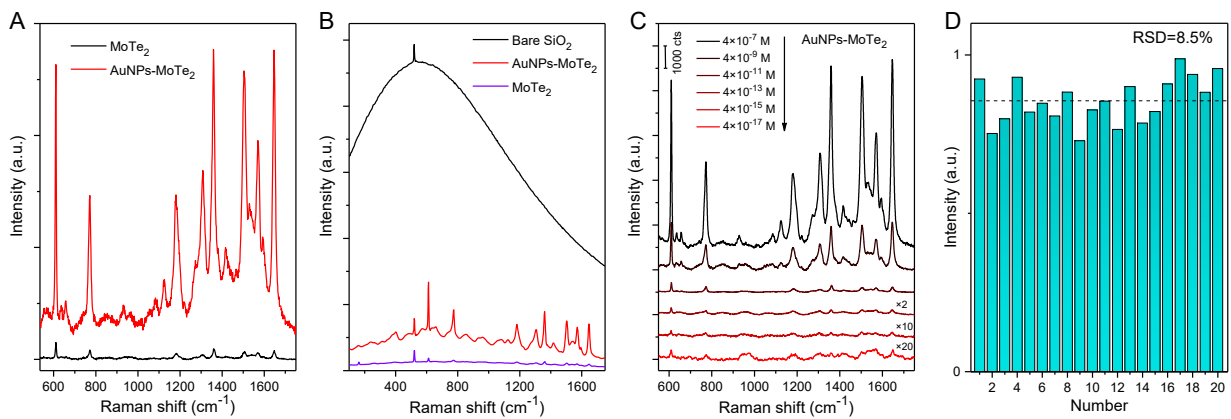

**Figure 3. SERS performance and FL quenching.**

(A) SERS signals of R6G ($4\times10^{-7}$ M) on bare MoTe$_2$ and AuNPs–MoTe$_2$ hybrid structure.

(B) Raman–FL spectra of R6G on bare SiO$_2$, MoTe$_2$ and AuNPs–MoTe$_2$ hybrid structure.

(C) SERS signals collected on AuNPs–MoTe$_2$ with various concentrations of R6G coating.

(D) Histograms of the SERS intensities of R6G 1646 cm$^{-1}$ feature peak on the AuNPs–MoTe$_2$ hybrid with R6G coating ($4\times10^{-7}$ M). The derived relative standard deviation (RSD) is 8.5%.

We have also investigated the influence of Raman excitation wavelength on the SERS sensitivity. As shown in **Figure S7**, the dye Raman features exhibit larger enhancement when the excitation is resonant with the probes, under which condition the charge transfer resonance can be maximized. Nevertheless, the dyes coated on AuNPs–MoTe$_2$ show acceptable SERS sensitivities with non-resonance excitation wavelengths, due to the incorporation of EM. On the other hand, the SERS effect can be barely observed on substrates based solely on CM when the excitation does not resonate with the probes.[11,36]

By taking the advantages of both plasmonic localized field enhancement and chemical enhancement, the novel AuNPs–MoTe$_2$ hybrid is expected to detect analyte at ultralow concentrations. **Figure 3C** presents the SERS spectra of R6G coated on AuNPs–MoTe$_2$ with various concentrations from $4\times10^{-7}$ to $4\times10^{-17}$ M. Distinct Raman fingerprint peaks of R6G are clearly detected. At ultralow R6G concentration of $4\times10^{-17}$ M, the SERS signals are still detectable, giving rise to the capability of detecting trace amount of analytes on the AuNPs–MoTe$_2$. This ideal limit of detection (LoD) is four orders of magnitude improved in comparison with the individual 1T′ MoTe$_2$ substrates,[11] and far surpasses the previously reported results on metal nanostructures hybridized with graphene, hBN, MoS$_2$ and other 2D materials, whose fabrication processes are much more complicated, costly and time-consuming.[16,17,37,38] Detailed comparison between the present work and other plasmonic–2D material based SERS substrates is listed in **Table S1**. The intensity of R6G 609 cm$^{-1}$ feature peak and concentration of coated R6G shows a linear relationship in log scale (**Figure S8**), suggesting that the AuNPs–MoTe$_2$ platform is promising for quantitative molecular detection.

The SERS intensities collected on AuNPs–MoTe$_2$ are uniformly distributed, as evidenced by the Raman intensity map of R6G 609 cm$^{-1}$ peak in **Figure S9A**. The SiO$_2$ areas shows no appreciable R6G Raman signals, since the AuNPs cannot be formed on bare SiO$_2$. SERS spectra collected from 20 randomly selected spots on AuNPs–MoTe$_2$ are shown in **Figure S9B**, and the corresponding histograms of the R6G Raman intensity depicted in **Figure 3D** reveal a small relative standard deviation of 8.5% in the SERS signals. This high homogeneity of the SERS signals is attributed to the atomic flat surface of the 2D MoTe$_2$ which uniformly adsorbs analytes, and the uniform morphology of the in-situ self-assembled AuNPs. This feature gives rise to the reproducible and quantitative molecular detection on the hybrid SERS platform, which is of technical importance in real-world applications. The photobleaching effect of dyes on metallic nanostructures is another factor that leads to the low reproducibility in conventional SERS detections. Even for some semiconductor SERS substrates, the photocatalytic degradation of the analyte is non-negligible.[39] The AuNPs–MoTe$_2$ hybrid, fortunately, can greatly eliminate the unwanted degradation of probes upon illumination. To conduct a comparative investigation of the photo stability, a control substrate of AuNPs on $SiO_2$/Si was fabricated by depositing 10-nm gold film on $SiO_2$ followed by vacuum annealing at 300 °C. As shown in **Figure 4**, the R6G Raman intensities on AuNPs–$MoTe_2$ mostly remain stable with the prolonged laser illumination, while those on the conventional substrate (AuNPs) present a significant decrease with increasing acquisition time. The highly improved stability against illumination on AuNPs–$MoTe_2$ can be attributed to the large charge transfer between the analytes and $MoTe_2$ that promotes the relaxation of analyte excitation states and dissipation of hot electrons.[40]

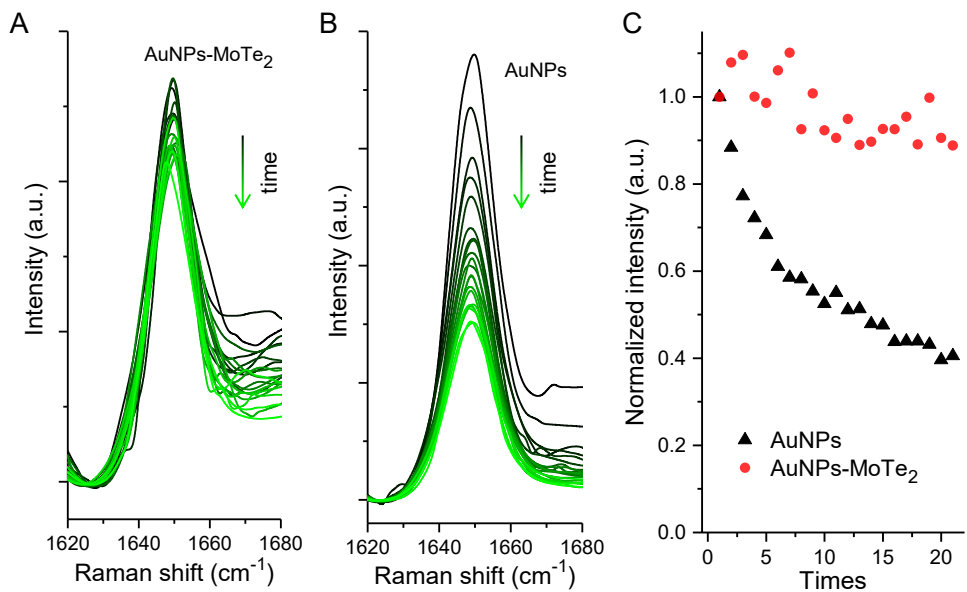

**Figure 4. SERS stability under illumination.**

(A and B) SERS signals of R6G 1646 $cm^{-1}$ feature peak on AuNPs–$MoTe_2$ hybrid structure (A) and bare AuNPs (B) under continuous laser illumination. Each spectrum was collected with 10 s acquisition time and 10 s interval before the next measurement.

(C) SERS intensities of R6G 1646 $cm^{-1}$ peak with increasing number of times of spectral acquisition.

The AuNPs–$MoTe_2$ hybrid structure is applicable for various analytical applications, such as detecting risky chemicals in food. We have tested trace amounts of Sudan III and crystal violet (CV) dyes on the hybrid SERS platform, which are category 3 carcinogen classified by IARC and harmful fish drug, respectively. The SERS signals of Sudan III and CV coated on AuNPs–MoTe$_2$ are plotted in **Figure S10A and S10B**. The analyte fingerprint peaks are clearly detected, even at an ultralow analyte concentration of $4\times10^{-15}$ M, showing the potential of AuNPs–MoTe$_2$ hybrid in food safety and environmental pollution monitoring with high sensitivity. The feature peak intensities of the analytes as functions of the analyte concentrations are given in **Figure S10C and S10D**.

**Flexible SERS tapes for real-world applications**

To realize real-time, in-situ and reliable SERS detection for daily-life applications, we have fabricated flexible and transparent SERS tapes based on AuNPs–MoTe$_2$ that can be applied on arbitrary surfaces. As shown in **Figure 5A**, a large-area continuous thin film of 1T′ MoTe$_2$ was synthesized on SiO$_2$/Si by tellurization of 2-nm pre-deposited Mo film (Raman spectrum and intensity map in **Figure S11** indicate the high quality and uniformity of the MoTe$_2$ film).[41] The MoTe$_2$ film was then transferred onto a polydimethylsiloxane (PDMS) stamp, and the AuNPs/MoTe$_2$/PDMS SERS tape was obtained after HAuCl$_4$ treatment (**Figure 5B**). Detailed fabrication process of the SERS tape is available in Experimental Procedures. Here we show two typical application scenarios. As shown in **Figure 5C**, the SERS tape can float on aqueous solution of analyte ($10^{-6}$ M R6G was used as an example), and the analyte Raman signals can be clearly detected with the SERS tape. **Figure 5D** is another application that the SERS tape sticks onto an apple surface. The apple was immersed in Sudan III solution ($10^{-6}$ M) for 20 min before the adhesion of SERS tape. Distinct Raman signals from the analyte molecules adsorbed on the apple surface is in-situ observed on the SERS tape covered area, while no analyte signal is collected on the area without SERS tape. The detection sensitivity of the SERS tape is expected to be improved by optimizing the growth process to obtain MoTe$_2$ continuous film with higher crystallinity, as grain boundaries would impede the charge transfer.

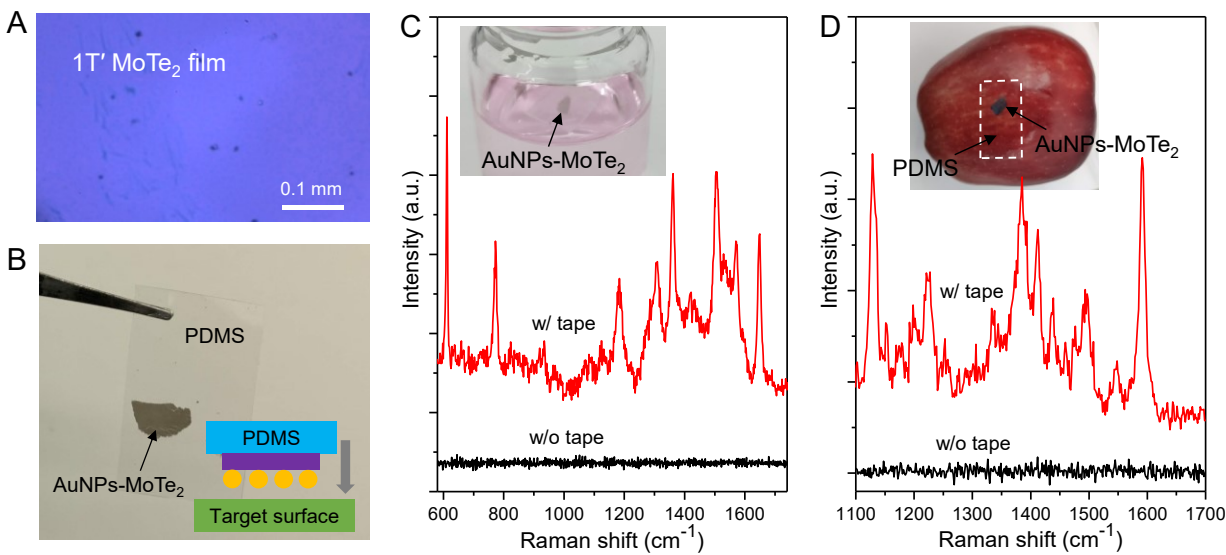

**Figure 5. Flexible SERS tape for practical applications.**

(A) Optical image of the large-area continuous thin film of 1T′ MoTe$_2$.

(B) Image of the flexible AuNPs/MoTe$_2$/PDMS SERS tape. Inset is the schematic showing that the SERS tape can be applied onto target surfaces for various applications.

(C) SERS detection with and without SERS tape floating on the R6G aqueous solution.

(D) SERS detection with and without SERS tape stuck on the apple surface with adsorbed Sudan III dye.

## DISCUSSION

We have demonstrated that uniform and high-density AuNPs can be in-situ prepared on 2D 1T′ MoTe$_2$ via a facile and fast chemical treatment without design constraints, spontaneously forming a hybrid structure for SERS. The high surface activity of the 1T′ phase not only is crucial for the formation of AuNPs, but also leads to the strongly interacted analyte–MoTe$_2$ interface. The AuNPs–MoTe$_2$ SERS platform generates coupling effects of plasmonic enhancement and chemical enhancement, rendering the ultrasensitive molecular detection. Additionally, the hybrid structure delivers uniform, reproducible and clean SERS signals, and the SERS signal intensities remain stable under prolonged laser illumination. Our work demonstrates that the easily fabricated hybrid SERS substrate is promising in practical applications such as food safety control, and sheds light on the understanding of electromagnetic and chemical enhancements in SERS.

**EXPERIMENTAL PROCEDURES**

**Material growth and characterization**

MoTe$_2$ atomic layers were synthesized in a home-built CVD system using space-confined strategy.[42] 20 mg ammonium molybdate tetrahydrate was placed in a quartz boat as the precursor. 2 mg NaCl was added in the same boat as the growth promoter.[43] SiO$_2$/Si substrate cleaned with acetone sonication was put on the boat with a face-to-face configuration. Another boat containing 50 mg tellurium powder was placed in the upstream region of the CVD chamber with a close distance to the growth substrate (1 cm). The growth was conducted at atmospheric pressure with Ar/H$_2$ (250/20 sccm) carrier gas environment. The furnace was heated up to 810 °C in 15 min and kept for 10 min for the growth of 1T′ MoTe$_2$. The growth temperature was decreased to 650 °C to produce 2H MoTe$_2$. In the main text, the MoTe$_2$ is of 1T′ phase unless specified. SEM images were obtained by SUPRA 60 with 10 kV accelerating voltage, and TEM images were carried out by Titan3 G2 60-300, FEI. AFM (NTEGRA, NT-MDT) was used to characterize the surface morphology. XPS measurements were conducted by ESCALAB 250Xi. Raman/FL spectra were conducted with Horiba LabRAM HR Evolution Raman system using 532 nm excitation unless specified.

**FDTD simulation**

A commercial FDTD simulation package (FDTD solutions, Lumerical Inc.) was used with 3D simulation for calculating the near field profile and scattering/absorption spectrum of the AuNPs/1T′ MoTe$_2$ hybrid structure. Two gold nanosphere of diameter 20 nm with a gap of 3 nm were set on 1T′ MoTe$_2$ with a thickness of 3 nm, the structure is on the SiO$_2$ substrate and the simulation background is air. A normal incident total field-scattering field (TFSF) source was used with polarization along the line of the two nanospheres centers. The anti-symmetric and symmetric boundary conditions were accordingly used for reducing the simulation time and resource requirements, the others were set to be perfectly matched layers. The mesh sizes for the regions of the two nanospheres and 1T′ MoTe$_2$ were set to be 0.5 and 0.25 nm, respectively. Two optical power analysis groups inside and outside the TFSF source region were used for calculating the absorption and scattering spectrum.

**DFT calculation**

DFT calculation were performed using the Perdew−Burke−Ernzerhof (PBE) functional within the generalized gradient approximation (GGA) used to describe the exchange−correlation interactions. And projector-augmented wave (PAW) pseudopotentials were used to describe the core electrons. All calculations were done using VASP code.[44] The wave functions were expanded in plane waves up to a cutoff of 400 eV. The Brillouin zone integration was sampled by a 2 × 2 × 1 k mesh. The systems were simulated with a periodic boundary condition by placing a $[AuCl_4]^-$ ion on the surface of 72-atom 1T′ $MoTe_2$ (Mo:24 and Te:48) and 75-atom 2H $MoTe_2$ (Mo:25 and Te:50), respectively. We performed full geometry optimizations until the residual forces were less than 0.01 eV/Å. To compute the reaction barriers, we used the NEB method with the 'climbing image' algorithm.[45] Activation barriers obtained with the NEB method refer to a temperature of 0 K. The reaction pathway by means of NEB calculations was based on the constrained geometry optimizations. Therefore, all intermediate states were identified by a series of constrained geometry optimizations. Once a constraint is defined (e.g., the forces between atoms), a geometry optimization is performed to force the constraint to preserve a given value.

**SERS measurements**

Sequential dilution process was performed for obtaining aqueous solution of the analytes (R6G, CV and Sudan III) with various concentrations. The SERS substrates were immersed into the analyte solution for 20 min for the coating of analyte molecules. SERS spectra were collected using a Horiba LabRAM HR Evolution Raman system with 532 nm excitation laser (unless specified) and 100× objective lens. The laser spot size is around 1 μm. For each spectral acquisition, the laser power was 3.2 mW and the integration time was 2 s (unless specified). For the SERS measurement with analyte concentration lower than $10^{-14}$ M, the signal intensities were averaged from five individual acquisitions to enhance the signal-to-noise ratio. Baseline correction was performed in the SERS spectra expect for the Raman–FL spectra in Figure 3B.

**Flexible SERS tape fabrication**

Mo film of 2 nm thickness was deposited on SiO$_2$/Si substrate via electron beam evaporation. The Mo film was then tellurized into 1T′ MoTe$_2$ continuous film in the CVD furnace at 810 °C. After the growth, the SiO$_2$ substrate was etched in HF aqueous solution (10%), and the 1T′ MoTe$_2$ film was then transferred onto a PDMS film (Gel-Pak). Note that supporting polymer such as poly(methyl methacrylate) was not applied during the transfer since the 1T′ MoTe$_2$ film is of excellent robustness. After gentle drying and baking (80 °C for 10 min) process, the MoTe$_2$/PDMS stack was immersed in the HAuCl$_4$ solution (1 mM) for 30 s for the formation of dense AuNPs on the MoTe$_2$ surface. Thus, the fabricated AuNPs/MoTe$_2$/PDMS flexible SERS tape was ready to apply on target surfaces for various applications.

**SUPPLEMENTAL INFORMATION**

Supplemental information can be found online.

**ACKNOWLEDGMENTS**

The work is in part supported by Research Grants Council of Hong Kong (Grant No. AoE/P-02/12), CUHK Group Research Scheme, ITS/390/18 and Research Talent Hub Scheme for ITF project by Innovation and Technology Commission, Hong Kong SAR Government, National Natural Science Foundation of China (Grant Nos. 62005051, 51802360, 61975036 and 61804067), National Natural Science Foundation of Guangdong for Distinguished Young Scholars (Grant No. 2018B030306043), Pearl River Talent Plan (Grant No. 2019QN01C109), Science and Technology

Program of Guangzhou (Grant No. 201904010449), and Key cultivation program for young teachers of Sun Yat-sen University (Grant No. 20lgzd13).

AUTHOR CONTRIBUTIONS

L.T. conceived the project with the supervision from J.B.X. and K.C. L.T. and K.C. synthesized and characterized the materials with assistance from Y.Zhou, H.L., X.Wang, and R.Z. L.T. performed the Raman experiments with assistance from K.C., X.H., and Y.Zhao. K.C. and Z.L. conducted the theoretical modeling and calculations. L.T., K.C. and J.B.X. wrote the paper. All authors discussed the results and commented on the manuscript.

DECLARATION OF INTERESTS

The authors declare no competing interests.